\documentclass[doublecol]{epl2} 

\usepackage{graphicx} 
\usepackage{amssymb} 
\usepackage{amsmath} 
\usepackage{verbatim}
\newcommand{\nn}{\nonumber} 
\newcommand{\veps}{\varepsilon}

\renewcommand{\eqref}[1]{Eq. (\ref{#1})} 
\newcommand{\figref}[1]{Fig. \ref{#1}}

\title{Attraction Between Like-Charge Surfaces in Polar Mixtures} 

\author{S. Samin\inst{1} \and Y. Tsori\inst{1}}
\shortauthor{S. Samin \etal}

\institute{                    
  \inst{1} Department of Chemical Engineering and The Ilse Katz Institute for Nanoscale
Science and Technology, Ben-Gurion University of the Negev, 84105 Beer-Sheva, Israel.}


\abstract{
We examine the force between two charged surfaces immersed in aqueous
mixtures having a coexistence curve. For a homogeneous water-poor phase, as the distance
between the surfaces is
decreased, a water-rich phase condenses at a distance $D_t$ in the range
$1-100$nm. At this distance the osmotic pressure can become negative leading to a
long-range attraction between the surfaces. The osmotic pressure vanishes at a distance
$D_e<D_t$, representing a very deep metastable or globally stable energetic state. We
give
analytical and numerical results for $D_t$ and $D_e$ on the Poisson-Boltzmann level.
}

\begin{document}

\maketitle

The forces between charged objects in electrolyte solutions are of fundamental importance
in biology and colloidal science. Within the Poisson-Boltzmann (PB) mean-field theory, in
a single pure solvent, the interaction between symmetrically charged colloids is always
repulsive \cite{israelachvili}. Experiments, on the other hand, have shown that highly
charged colloids
can attract each other when multivalent ions are present. This discrepancy has
been
explained theoretically by counterion correlations \cite{Kjellander1986},
fluctuation-induced forces \cite{pincus2000} and other non-electrostatic interactions
\cite{andelman2000}. More recently, the PB theory was generalized by strong coupling
theory \cite{Netz2000,naji2010}, which predicts an attraction for large values of the so-called
coupling parameter \cite{Netz2001.2,Netz2001.1}.

In this Letter we use the PB theory to show that strong attractive forces appear between
similarly charged colloids in {\it mixtures}. In mixtures one must also take into
account that the medium itself becomes inhomogeneous due to dielectrophoretic
and solvation-related forces.
The preferential solvation energy of ions in one of the solvents \cite{israelachvili,marcus1988} is
appreciable and can even be much larger than the thermal energy
\cite{Osakai1998,Hung1980,Marcus1983}. For a recent review on ion-specific solvation effects within the PB theory
 see \cite{benyaakov2011}. Previous works on preferential solvation in binary
mixtures looked at the phase behavior in the bulk \cite{onuki:021506,PhysRevE.82.051501},
surface tension \cite{onuki2008}, and the interaction between surfaces but not in
immiscible
liquids \cite{andelman_jpcb2009}. As is shown below, in partially miscible mixtures, the
behavior is qualitatively different and strong forces occur. In colloidal suspensions 
these forces can have an important role not studied before 
\cite{Beysens1985,Beysens1998,Bechinger2008,nellen2011,Bonn2009}.

We consider two positively charged colloids in a mixture of polar solvents.
The colloids are modeled as flat surfaces located at $z=\pm D/2$ and uniformly
charged with a charge density $e\sigma$ per unit area, where $e$ is the
elementary charge.
A small amount of monovalent ions and weakly charged surfaces are assumed
and therefore the coupling parameter is small and the PB theory is applicable
\cite{naji2010}. 
The partially miscible solvents have a coexistence
curve below the mixture critical temperature $T_c$ in the absence of ions. This
coexistence curve is further
modified by the presence of ions \cite{PhysRevE.82.051501}. 

The grand potential density is given by \cite{onuki:021506,tsori_pnas_2007} 
\begin{align}
\frac{\omega}{T}&=f_{b}(\phi)+\frac{C}{2}|\nabla \phi|^2 +\frac{1}{2T} \veps(\phi)(\nabla
\psi)^2 \nn \\ &+n^+\left(\log (v_0n^+)-1\right)+n^-\left(\log (v_0n^-)-1\right) \nn\\
&-(\Delta u^+n^++\Delta u^-n^-)\phi-\lambda^+n^+-\lambda^-n^--\mu \phi 
\label{eq:fmions}
\end{align}
Here the Boltzmann constant is set to unity, $T$ is the thermal energy and $C$ is a
positive constant. $\phi$ is the volume fraction of water ($0\leq\phi\leq 1$); 
far away from the surfaces the mixture is homogeneous and water-poor 
with composition $\phi_0<1/2$ and ion densities $n_0$.
We use a regular solution form for the free energy of a binary
mixture $v_0f_b=\phi\log(\phi)+(1-\phi)\log(1-\phi)+\chi\phi(1-\phi)$
\cite{safran}, where $\chi\sim 1/T$ is the Flory parameter and $v_0=a^3$ is
the molecular volume of both liquids. The third term in \eqref{eq:fmions} is the
electrostatic energy of the mixture, where $\psi$ is the electrostatic potential. The
mixture dielectric constant depends on the composition through a linear
relation: $\veps(\phi)=\veps_c+(\veps_w-\veps_c)\phi$, where $\veps_c$ and $\veps_w$ are
the pure cosolvent and water dielectric constants, respectively. The second line of
\eqref{eq:fmions} is the ideal gas entropy of point-like ions, where $n^+$ and $n^-$ are
the positive and negative ion density, respectively. The first term on the third line is
the bilinear solubility interaction of the ions and the solvent: the parameters $\Delta
u^+$ and $\Delta u^-$ measure the affinity of the positive and negative ions toward the
water environment, respectively. We use the common case where both ions are hydrophilic
and assume the symmetric interaction $\Delta u^+=\Delta u^-=\Delta u>0$. Lastly,
$\lambda^\pm$ and $\mu$ are the Lagrange multipliers (chemical potentials) of the positive
and negative ions and water composition, respectively.

The ion densities obey the Boltzmann distribution
\begin{align} \label{eq:npm}
n^\pm&=v_0^{-1}{\rm e}^{\lambda^\pm}{\rm e}^{\mp e \psi/T+\Delta u\phi} 
\end{align}
In a
salt reservoir we have $\lambda^\pm=\log(v_0 n_0)-\Delta u \phi_0$. Alternatively, when
the mixture contains only negative counterions, $\lambda^-$ is determined
self-consistently by charge conservation: $\int n^- {\rm d}z =2\sigma$.
 
The electrostatic potential is determined by the Poisson equation 
\begin{align}\label{eq:laplace}
\nabla \cdot
(\veps(\phi)\nabla \psi)=-e(n^+-n^-)
\end{align}
supplemented by the boundary condition on the surfaces
$-\mathbf{n}\cdot\nabla\psi(\pm D/2)=e\sigma/\veps(\phi)$, where $\mathbf{n}$ is the
outward unit normal to the surface. Finally, the Euler-Lagrange equation for $\phi$ reads
\begin{align} 
\label{eq:comp} C\nabla^2\phi=\frac{\partial f_{b}}{\partial \phi}
-\frac{1}{2T} \frac{d\veps}{d\phi}(\nabla\psi)^2-\Delta u(n^++n^-)-\mu 
\end{align}
In
order to isolate solvation and electrostatic effects, we assume zero short-range chemical
or long-range van der Waals interactions with the surfaces, leading to the boundary
condition $-\mathbf{n}\cdot\nabla\phi(z=\pm D/2)=0$.

The net force exerted on the surfaces by the liquid is given by the osmotic pressure
$\Pi=P_{zz}-P_b$, where $P_b$ is the bulk pressure: $P_b/T=\phi_0\partial
f_b(\phi_0)/\partial \phi-f_b(\phi_0) +2n_0\left(1-\Delta u \phi_0\right)$. $-P_{zz}$
is $zz$ component of the Maxwell stress tensor \cite{landau2,andelman_jpcb2009}:
\begin{align} 
\label{eq:pi} &\frac{P_{zz}}{T}=\frac{C}{2}\left(\frac{d\phi}{dz}\right)^2
-C\phi\frac{d^2\phi}{dz^2}+\phi\frac{\partial f_{b}}{\partial \phi}-f_{b} \nn \\ 
&+(1 -
\Delta u\phi) (n^+ + n^-)-\frac{1}{2T}\left( \phi \frac{d\veps}{d\phi} +
\veps\right)\left(\frac{d\psi}{dz}\right)^2 
\end{align}
$P_{zz}$ is independent of $z$
and can be calculated at the midplane ($z=0$) where by symmetry $d\phi/dz=d\psi/dz=0$.
Multiplying \eqref{eq:comp} by $\phi$ and inserting into \eqref{eq:pi} we obtain:
\begin{align} 
\label{eq:pi_mid} \Pi=Tn_m-T\omega_{b}(\phi_m)-P_b 
\end{align}
where $\omega_b=f_b-\mu\phi$ and $\phi_m$ and $n_m$ are the composition and total ion
density at the midplane, respectively.

\begin{figure}[!tb] 
\includegraphics[width=8.5cm,viewport=95 280 515 510]{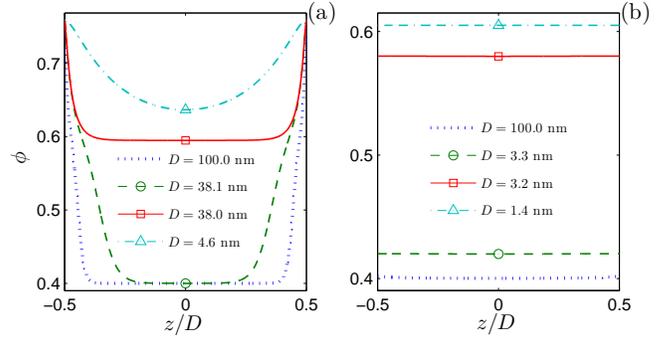}
\caption{(a) Composition profiles $\phi(z)$ in the strong screening limit for $D<D_t$
(dotted line), just before and after $D_t$ (dashed and solid lines) and at $D_e$ (dash-dot
line). Here $n_0=0.1$M, $\Delta u=2$ and $\sigma=1$nm$^{-2}$. (b) The same in the ideal
gas regime with no salt, $\Delta u=11$ and $\sigma=0.01$nm$^{-2}$. Here and, unless stated otherwise, in all other
figures we took for the mixture $\phi_0=0.4$, $T/T_c=0.9915$ and $C=\chi/a$ \cite{safran}.
As an approximation of a water--1-butanol mixture we used $T_c=398$K, $v_0=3\times
10^{-29}$m$^3$, $\veps_a=17.8$ and $\veps_w=80$.} 
\label{fig1} 
\end{figure}

\begin{figure}[!tb] 
\includegraphics[width=8cm]{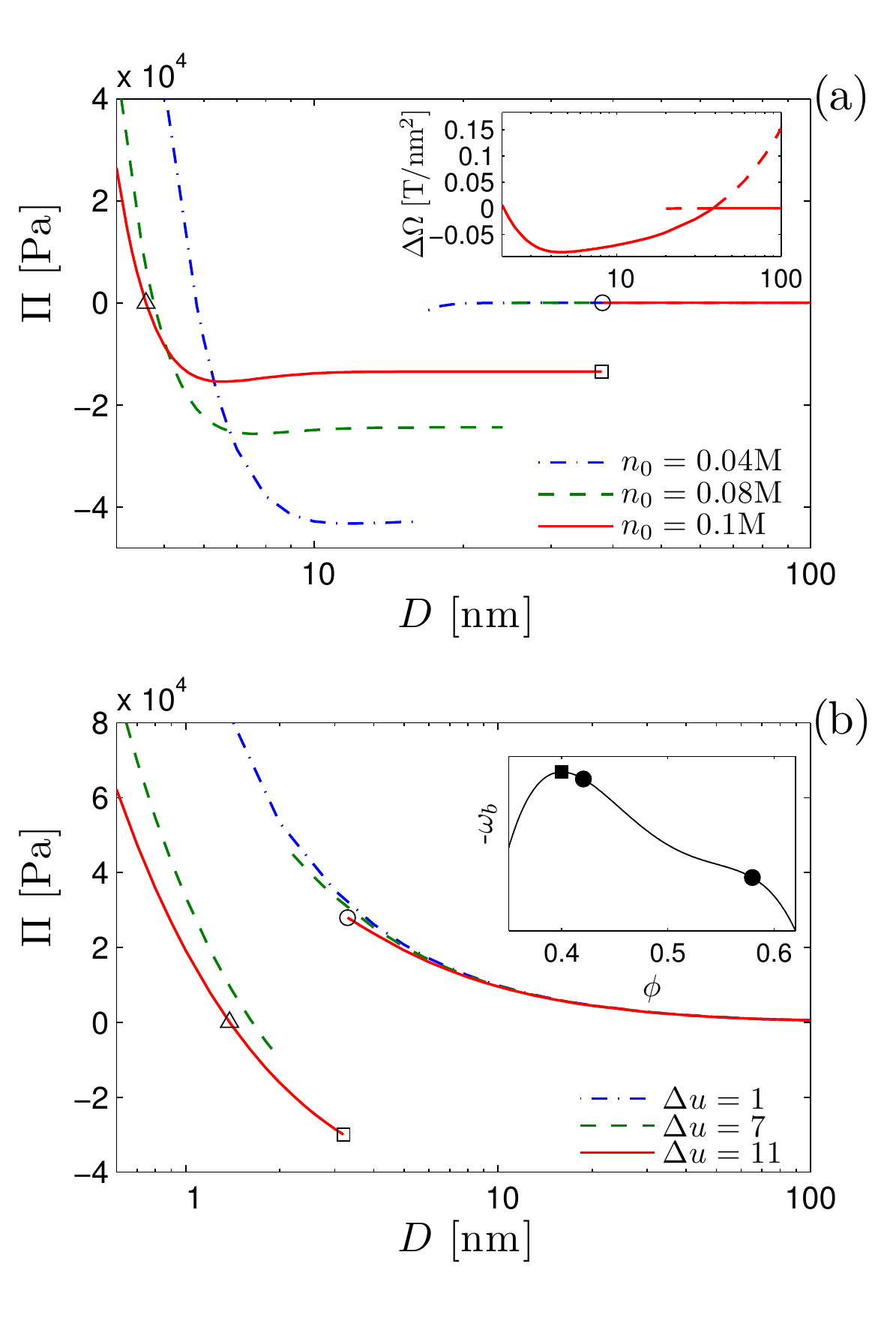}
\caption{(a)
Osmotic pressure $\Pi$ vs surface separation $D$ in the strong screening limit for
different values of $n_0$ [other parameters as in \figref{fig1}(a)]. Inset: excess
grand potential for $n_0=0.1$M. solid (dashed) line corresponds to stable (metastable)
solutions. (b) $\Pi$ in the
ideal gas regime for different values of $\Delta u$ [other parameters as in
\figref{fig1}(b)]. Inset: the function $-\omega_{b}(\phi)$. Filled square and circle
markers
correspond to $\phi_0$ and the binodal compositions, respectively. 
In (a) and (b) open markers on solid lines correspond to distances $D$ with the same
symbols in \figref{fig1} (a) and (b).
} 
\label{fig2}
\end{figure}

\begin{figure}[!tb] 
\includegraphics[width=8cm,viewport=4 0 335 270,clip]{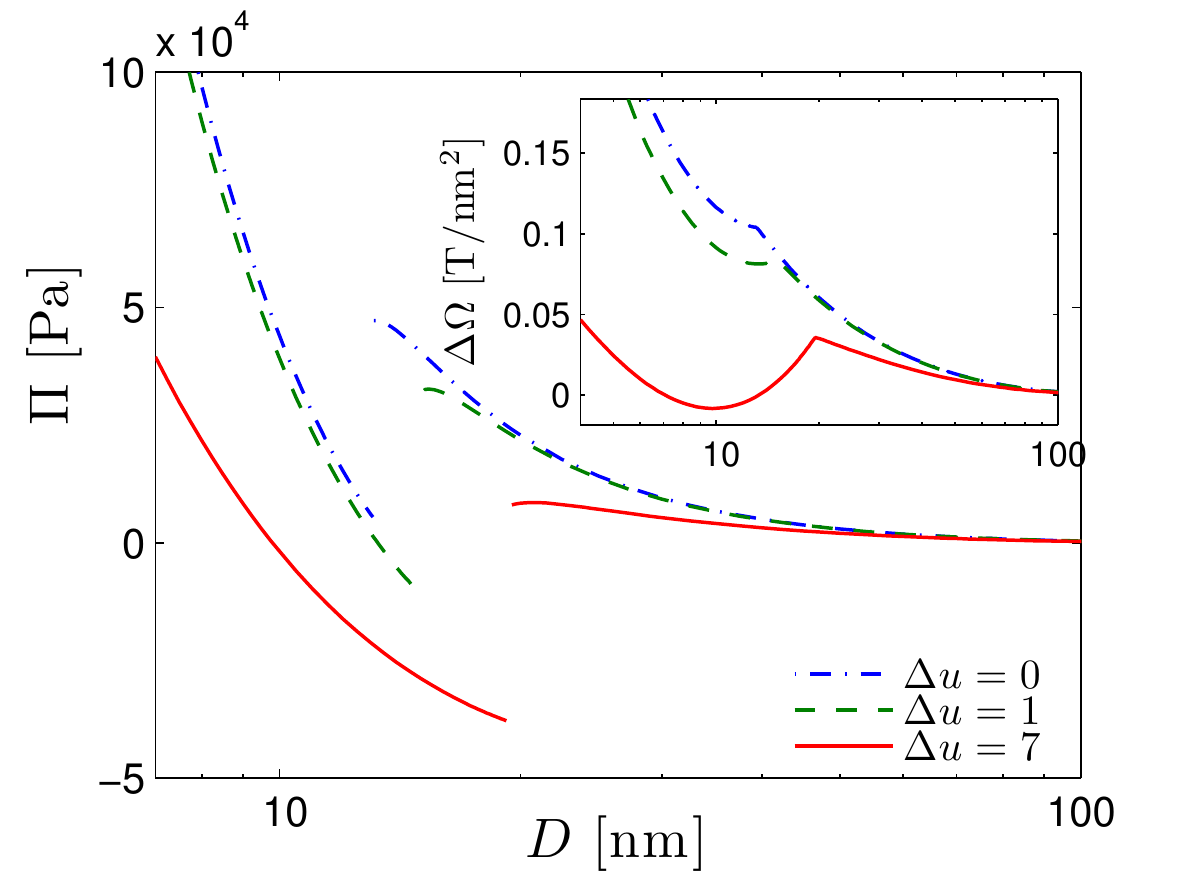}
\caption{Osmotic pressure in the weak screening regime with $n_0=10^{-4}$M and 
$\sigma=1$nm$^{-2}$ for three values of $\Delta u$. Inset: the corresponding excess
grand potential $\Delta \Omega$.} 
\label{fig3} 
\end{figure}

In aqueous mixtures, water is drawn towards the walls by field gradients
(dielectrophoretic) and solvent-induced
(electrophoretic) forces \cite{tsori_pnas_2007, andelman_jpcb2009}. 
These forces are interdependent as is evident from Eqs. (\ref{eq:laplace}) and
(\ref{eq:comp}) . From the solution of
the governing equations we find two possible scenarios, distinguished according to the
composition profiles at infinite separation. In the first scenario, when the surfaces are
far apart, $\phi(z)$ near each surface is composed of three layers: a thin water-rich
layer
($\phi>1/2$) of width $t$ at the surfaces, a second layer with width of the bulk
correlation length
$\xi$ adjacent to it, where the composition $\phi \gtrsim \phi_0$
decays to the bulk value, and a third ``bulk'' layer where $\phi\approx\phi_0$.

As the surface separation decreases down to $D=D_t$, the general features of this profile
do not
change. In particular, only the width of the $\phi \approx \phi_0$ region changes 
if screening is strong, that is, when $l_D \ll D$ where
$l_D\sim n_0^{-1/2}$ is the Debye length calculated at $\veps=\veps(\phi_0)$. The dashed
curve in
\figref{fig1}(a) shows $\phi(z)$ just
before $D_t$ for this case.

In the second scenario, dielectrophoretic and solvation forces are not strong enough and
at large surface separations $\phi$ is always close to the bulk value $\phi_0$. As $D$
decreases from infinity the water composition between the surfaces increases continuously.
In particular, in the ideal gas regime, defined by $b\gg D$, where $b \propto \sigma^{-1}$
is the
Gouy-Chapman length, $\phi(z)$ is nearly uniform. \figref{fig1}(b) shows such profiles for
a mixture containing only counterions. Here, the ion profile $n(z)$ and electric field 
are nearly uniform leading to small composition gradients. 

In both scenarios, when $D<D_t$ the whole space between the surfaces becomes rich in
water (see the solid curves in \figref{fig1}). The transition occurs at a 
distance in the range
$D_t=1$--$100$nm depending on the parameter values, and is accompanied by
a decrease
in the osmotic pressure. This can lead to a negative osmotic pressure, such that a
surface separation $D_e<D_t$ exists at which $\Pi=0$. $\phi(z)$ at $D=D_e$ is shown in the
dash-dot curves in \figref{fig1}. 

The excess grand
potential per unit area relative to infinite separation is $\Delta
\Omega(D)=-\int_{\infty}^{D} \Pi(D') {\rm d} D'$.  $\Delta \Omega(D)$ has a cusp at
$D=D_t$ and this
results in a discontinuity of the pressure. Close to
$D_t$, both a water-poor and a water-rich profiles are solutions of the
Euler-Lagrange equations but only one of these is a stable solution. At $D>D_t$ the
water-poor phase is stable, at $D<D_t$ the water-rich phase is stable, and  
at $D=D_t$ the grand potentials of the two phases are equal.
Furthermore, $\Delta \Omega$ has a minimum at $D_e$ ($D_e<D_t$) corresponding to
mechanical equilibrium.
An example of $\Delta\Omega(D)$ is shown in the inset of \figref{fig2}(a), where the
dashed line is $\Delta\Omega$ of metastable solutions.

The transition to a water-rich phase has a different physical origin in the two limiting
cases. In the strong screening regime, the transition is promoted by the energy gained 
when the
interface vanishes, as in capillary condensation. The thin
water-rich layer near the walls remains nearly independent on $D$.
In the ideal gas regime, 
on the other hand, when
the composition approaches the coexistence composition, preferential solvation
promotes a transition to a water-rich phase. For
intermediate cases, both mechanisms play a role in the transition. 
These effects are enhanced but are not limited to the vicinity of the critical
temperature where the differences between the phases become smaller.

In \figref{fig2} we plot the osmotic pressure as a function of surface spacing; when
$\Pi>0$
the surfaces repel each other while for $\Pi<0$ they attract. 
$\Pi$ is discontinuous at $D=D_t$. 
In \figref{fig2}(a) we show $\Pi$ for different values of $n_0$ in
the strong screening limit. 
An increase in $n_0$ decreases $\Pi$ at large distances but 
increases it at small distances (entropy loss of the ions). 

The negative jump in $\Pi$ at $D=D_t$ increases with decreasing $n_0$. This can be explained
by the interplay between the first two terms in \eqref{eq:pi_mid}. At $D=D_t$ the increase
in $\phi_m$ dominates the decrease in $\Pi$, an effect more pronounced for smaller 
values of $n_0$
since the positive ideal gas term is proportional to $n_m \propto n_0$. Recall that in a
pure solvent only the ideal gas term exists and hence $\Pi$ is always positive.
Furthermore, this entropic repulsion will eventually dominate $\Pi$ leading to
$\Pi=0$ at a separation $D_e$.

In \figref{fig2}(b) we plot $\Pi$ for different values of $\Delta u$ in the ideal gas regime.
For $\Delta u =1$ the pressure is purely repulsive since preferential solvation is not
strong enough to induce a water-rich phase. For $\Delta u =7$ the interaction is repulsive
down to $D_t$ where $\Pi$ becomes negative. When $\Delta u$ increases to $11$ the
transition
is at a larger distance and to a more negative pressure. The inset of
\figref{fig2}(b) shows the function $-\omega_{b}(\phi)$, showing a decrease in pressure at
the transition [cf.
\eqref{eq:pi_mid}].

\figref{fig3} shows $\Pi$ in the weak screening regime where $l_D\approx D$
and for three different values of $\Delta u$.
When $\Delta u =0$, the
dielectrophoretic force alone can induce a water-rich phase albeit the pressure
is always repulsive. The pressure can become attractive for $\Delta u =1$ or $7$. 
Corresponding curves of $\Delta \Omega$ are plotted in
the inset of \figref{fig3}. These show a metastable minimum at a finite value of $D$,
$D=D_e$, for $\Delta u=1$
and
a global minimum for $\Delta u =7$. The depth of the minimum for $\Delta u =7$ is 
$\approx 440T$ for two colloids with an effective surface area of $100$nm$\times 100$nm.
 
\begin{figure}[!tb] 
\includegraphics[width=8cm,viewport=4 20 335 490,clip]{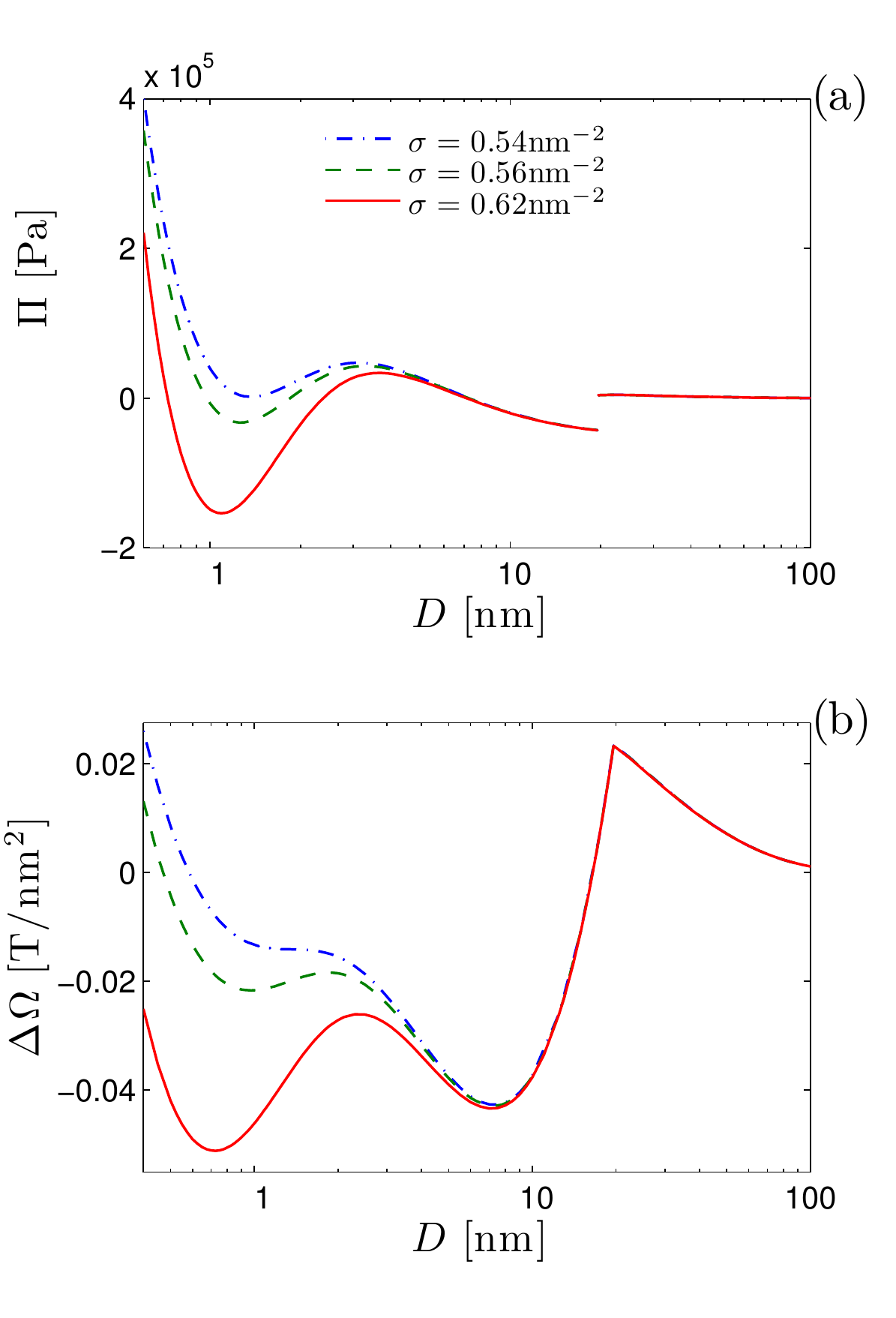}
\caption{(a) osmotic pressure for a large $\Delta u=10$, $n_0=10^{-4}$M and three values 
of $\sigma$. (b) the corresponding excess grand
potential. For $\sigma=0.54$nm$^{-2}$, $\Delta\Omega$ has only one minimum (dash-dot 
curve) at $D<D_t$. A second, metastable minimum appears for $\sigma=0.56$nm$^{-2}$ (dashed
curve), whereas for $\sigma=0.62$nm$^{-2}$ (solid curve) this minimum becomes globally
stable.} 
\label{fig4} 
\end{figure}

For $\Delta u =10$, a second metastable minimum can appear in the
curve $\Delta\Omega(D)$ at a smaller
value of $D$, see the dashed and dash-dot curves in \figref{fig4}. 
A similar value of $\Delta u$ is cited in the literature for mixtures of water and 1-butanol
containing NaCl at room temperature \cite{Marcus1983}. 
When $\sigma$ is
further increased (solid curves), this minimum can become globally stable. For large
enough $\sigma$, only the second minimum exists (not shown). In this large $\Delta u$
case, preferential solvation leads to the liquid between the surfaces being nearly pure
water; $\phi_m$ is close to unity and $\omega_b$ diverges.
\begin{figure}[!tb] 
\includegraphics[width=8.8cm,clip]{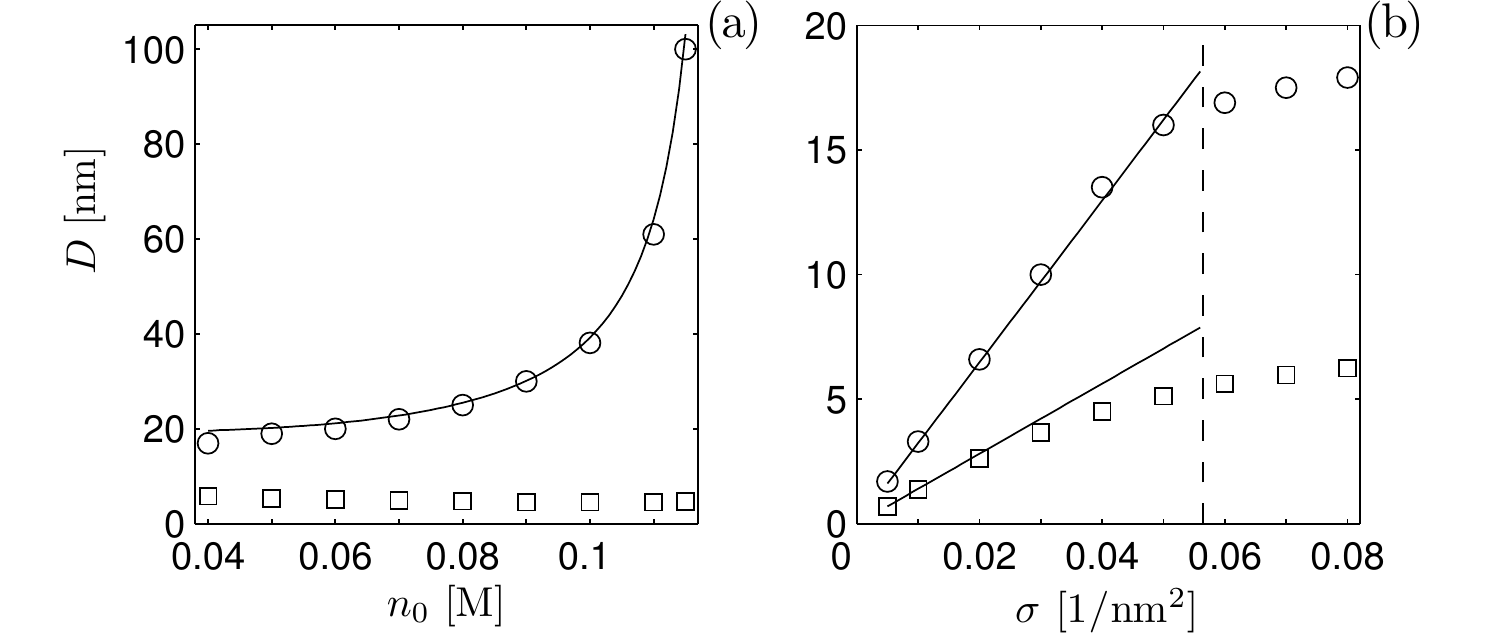}
\caption{Transition and equilibrium separations $D_t$ (circles) and $D_e$ (squares).
(a) The strong screening case vs $n_0$. (b) The ideal gas regime vs $\sigma$. Solid curves
are analytical expressions given in the text. In (a) we approximated
$\phi(z)$ by a linear decrease from $\phi_h$ to $\phi_0$ at the interface region.
In (b), for $\sigma$ beyond the dashed
curve, a water-rich layer near the surfaces exists even as $D \rightarrow \infty$.} 
\label{fig5}
\end{figure}
\begin{figure}[!tb] 
\includegraphics[width=8.8cm,]{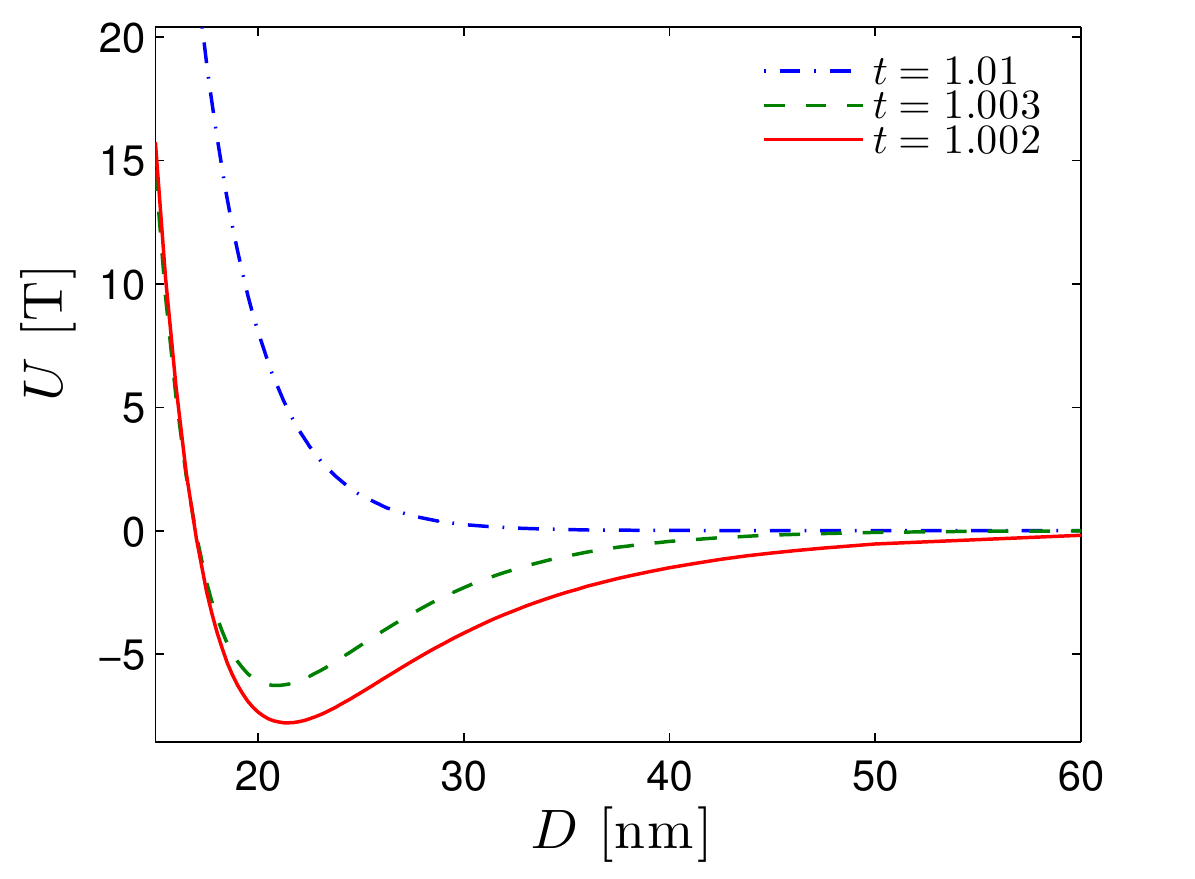}
\caption{Interaction potential for a mixture with a critical composition $\phi_0=1/2$ at three temperatures above $T_c$. For $t=1.01$ (dash-dot curve) the interaction is purely repulsive. When the temperature is decreased the interaction turns attractive (dashed curve), becoming stronger closer to $T_c$ (solid curve). Here $n_0=0.01$M, $\Delta u=4$ and $\sigma=1$nm$^{-2}$. The walls have a surface of $A=0.01\mu$m$^2$.} 
\label{fig6} 
\end{figure}

The distance $D_t$ can be obtained in the two limiting cases presented in
\figref{fig1}. In the strong screening limit we use continuity of $\Delta \Omega$ at $D_t$
to obtain \cite{Evans1986}: 
\begin{align} 
\label{eq:Dt_strong} 
D_t \simeq 2 t+2\frac{\xi P_b +
\int_{D/2-\xi-t}^{D/2-t} ~\omega(\phi(z)) {\rm d}z}{P_b-P_h} 
\end{align}
$P_h$, the pressure in the water-rich phase, is calculated from 
$P_h=T(n_h^++n_h^-)-T\omega_b(\phi_h)+P_b$, where 
$n^{\pm}_h=n_0{\rm e}^{\Delta u (\phi_h-\phi_0)}$ are the midplane values
and $\phi_h$ is obtained from 
\begin{align}
\label{eq:comp_sym} 
\frac{\partial f_{b}}{\partial \phi}(\phi_h)-2\Delta u n_0 {\rm
e}^{\Delta u (\phi_h-\phi_0)}-\mu_0=0 
\end{align}
where we used $\psi\approx 0$ at the midplane.

At $D=D_t$ the surfaces can be regarded as isolated and a zeroth-order approximation of
$\psi$ can be obtained using the well known result for a single surface
\cite{israelachvili}
with a homogeneous $\phi\approx\phi_h$ and a modified Debye length
$\left(\veps(\phi_h)T/\left(2e^2n_0 {\rm e}^{\Delta
u(\phi_h-\phi_0)}\right)\right)^{1/2}$. Since at $|z|<D/2-t$
$e\psi/T \ll \Delta
u(\phi_h-\phi_0)$, $t$ is obtained from the condition $e\psi/T =0.01 \Delta
u(\phi_h-\phi_0)$. In analogy with classical mean field theory \cite{safran}, the Landau
expansion of \eqref{eq:comp_sym} around $\phi_c=1/2$ gives for $\xi$ 
\begin{align} 
\xi \approx
v_0^{1/3}\frac{1}{\sqrt{1-\frac{T}{T_c}-2 \Delta u v_0 n_0 {\rm e}^{\Delta u (\frac12
-\phi_0)}\left(1+\frac{\Delta u}{4}\right)}}
\end{align}
The preferential solvation term in the root is comparable in 
magnitude to $1-T/T_c$. Thus, the bulk correlation length is modified
appreciably by the preferential solvation of the ions.

\figref{fig5}(a) shows the comparison of $D_t$ from \eqref{eq:Dt_strong} with numerical
results for different values of $n_0$. The agreement is quite good despite the
crude approximation. $D_e$ in \figref{fig5}(a) decreases slightly when $n_0$ increases,
its value being $\approx 5$nm.

In the ideal gas regime $\phi(z)\approx const.$ and the counter ions density $n^-$ is
uniform and equal to the average charge density \cite{safran}:
$\left<n^-\right>\simeq 2\sigma/D$. 
The composition equation reads
\begin{align} 
\label{eq:phim}
\frac{\partial f_{b}}{\partial \phi}-\frac{2\Delta u \sigma}{D}-\mu_0=0 ~.
\end{align}
Here preferential solvation merely shifts the chemical
potential of the mixture. Hence, $D_t$ occurs when $\phi$ is 
the binodal composition at which
$\partial f_{b}/\partial \phi=0$. Thus, 
\begin{align} 
\label{eq:Dt_low}
D_t\simeq -2\sigma\Delta u/\mu_0
\end{align}
which means that $D_t \rightarrow
\infty$ when the bulk composition approaches the binodal 
($\mu_0 \rightarrow 0$).

$D_e$ can be found by noting that $\Pi=0$ in \eqref{eq:pi_mid} gives
$n_e=\omega_{b}(\phi_{e})+P_b/T$, where $\phi_e$ and $n_e$ are the composition and
ion density at $D_e$, respectively. Inserting this into \eqref{eq:comp}
we obtain
for $\phi_e$ 
\begin{align} 
\frac{\partial \omega_{b}}{\partial \phi}(\phi_{e})=\Delta u
\left(\omega_{b}(\phi_{e})+P_b/T \right)
\end{align}
which is solved and 
then used to get $n_{e}$ and $D_e$ from the equations for $\Pi$ and $\left<n^-\right>$, 
respectively. 
Comparison of the formulae for $D_t$ and $D_e$
with the full numerical results are presented in \figref{fig5}(b). As expected, the
agreement is
good for small values of $\sigma$ where the ideal gas limit is valid. The dashed curve in
\figref{fig5}(b) marks the charge density above which a water-rich layer exists at
infinite separation and the above approximations no longer hold.

We find that at an attractive interaction is also possible at $T>T_c$. In this case, since there is no miscibility gap above $T_c$, the water-reach transition is missing and hence no discontinuity in the pressure is observed. Instead the pressure becomes smoothly attractive as the plate distance is reduced. \figref{fig6} shows the interaction potential $U=\Delta\Omega\times A$ for two plates in a critical mixture at three different temperatures. As the critical temperature is approached, a purely repulsive potential (dash-dot curve) becomes attractive (dashed curve). The interaction is more attractive closer to $T_c$ (solid curve). The interaction energy and length scales and the temperature trend shown in \figref{fig6} are similar to those observed in recent experiments on the salt-dependent interaction of a charged particle suspended in a critical binary liquid mixture near a charged wall \cite{nellen2011}. In light of this, we believe that the mechanism we describe may be important to capture correctly electrostatic effects in these experiments. Thus, it is possible to attribute some of the effects shown in these experiments to solvation related forces \cite{bier2011} in addition to the critical Casimir forces.

The qualitative difference in the interaction between charged 
colloids or macromolecules in mixtures compared to pure
solvents stems from the contribution to $\Pi$ of the second term in \eqref{eq:pi_mid}: $\omega_b(\phi_m)$ -- the midplane mixture grand potential. This term is absent in pure solvents. Thus, unlike similar ion induced phase transitions in pure solvents \cite{harries2006} we observe a jump to a \emph{negative} pressure (at $D_t\approx
1-100$nm). Moreover, the nontrivial dependence of $\omega_b(\phi_m)$ on the system parameters ($T$,$\phi_0$,$n_0$,$\sigma$) through the governing equations leads to qualitatively different behavior compared to usual condensation transitions due to surface fields \cite{Evans1986}. In addition the attractive force is in many cases strong and
long-range compared to the van der Waals force \cite{israelachvili} as is evident by the 
values of $\Pi$ at $D=D_t$ [\figref{fig2} and \figref{fig3}] and the 
energy minimum being deeper than $\sim 100 T$.

The mechanism we describe should be at
play in 
the aggregation of charged colloids in mixtures near the coexistence temperature
\cite{Beysens1985,Beysens1998}. We believe it is directly relevant to the
attraction seen between colloids
and surfaces in mixtures also far from $T_c$ and attributed to critical Casimir
forces
\cite{Bechinger2008,nellen2011,Bonn2009}.

\acknowledgments
We gratefully acknowledge numerous discussions with D. Andelman, C.
Bechinger, H. Diamant, J. Dietrich, L. Helden, O. Nelen, A. Onuki, P. Pincus, R.
Podgornik, S. Safran and M. Schick. This work was supported by the 
Israel Science Foundation under grant No.
11/10 and the European Research Council ``Starting Grant'' No. 259205.

\end{document}